\documentclass[conference,letter]{IEEEtran}
\IEEEoverridecommandlockouts
\usepackage{cite}
\usepackage{amsmath,amssymb,amsfonts}
\usepackage{algorithmic}
\usepackage{graphicx}
\usepackage{textcomp}
\usepackage{float}
\usepackage{xcolor}
\usepackage{booktabs}
\usepackage{url}
\usepackage{multirow}
\usepackage{balance}
\usepackage{color, colortbl}
\usepackage{placeins}
\usepackage{makecell}
\usepackage{flushend}
\usepackage{listings}
\usepackage{adjustbox}
\usepackage{array}

\newcolumntype{R}[2]{%
    >{\adjustbox{angle=#1,lap=\width-(#2)}\bgroup}%
    l%
    <{\egroup}%
}

\usepackage[left=2.56cm,right=2.56cm,top=0.75in]{geometry}

\definecolor{codegreen}{rgb}{0,0.6,0}
\definecolor{codegray}{rgb}{0.5,0.5,0.5}
\definecolor{codepurple}{rgb}{0.58,0,0.82}
\definecolor{backcolour}{rgb}{0.902,0.902,0.902}

\lstdefinestyle{mystyle}{
    backgroundcolor=\color{backcolour},   
    commentstyle=\color{codegreen},
    keywordstyle=\color{magenta},
    numberstyle=\tiny\color{codegray},
    stringstyle=\color{codepurple},
    basicstyle=\ttfamily\footnotesize,
    breakatwhitespace=false,         
    breaklines=true,                 
    captionpos=b,                    
    keepspaces=true,                 
    numbers=left,                    
    numbersep=5pt,
    xleftmargin=.12in,
    showspaces=false,                
    showstringspaces=false,
    showtabs=false,                  
    tabsize=2
}

\usepackage{subcaption}
\definecolor{LightGray}{gray}{0.9}

\def\BibTeX{{\rm B\kern-.05em{\sc i\kern-.025em b}\kern-.08em
    T\kern-.1667em\lower.7ex\hbox{E}\kern-.125emX}}
\begin{document}

\newcommand{\ie}{\textit{i}.\textit{e}.,\ }
\newcommand{\eg}{\textit{e}.\textit{g}.,\ }
\newcommand{\cf}{\textit{c}\textit{f}.\ }
\newcommand{\solution}{\textit{NOME}}
\newcommand\1{\textit{(i)}}
\newcommand\2{\textit{(ii)}}
\newcommand\3{\textit{(iii)}}
\newcommand\4{\textit{(iv)}}
\newcommand\5{\textit{(v)}}
\newcommand\6{\textit{(vi)}}
\newcommand\7{\textit{(vii)}}
\newcommand\8{\textit{(viii)}}
\newcommand\9{\textit{(ix)}}

\newcommand\itemA{\textit{(a)}}
\newcommand\itemB{\textit{(b)}}
\newcommand\itemC{\textit{(c)}}
\newcommand\itemD{\textit{(d)}}
\newcommand\itemE{\textit{(e)}}

\newcommand{\code}[1]{\texttt{#1}}

\title{Assessing SSL/TLS Certificate Centralization: Implications for Digital Sovereignty}

\author{\IEEEauthorblockN{Andrei Cordova Azevedo, Eder John Scheid, \\Muriel Figueredo Franco, Lisandro Zambenedetti Granville}
\IEEEauthorblockA{\\Institute of Informatics (INF)
\\Federal University of Rio Grande do Sul (UFRGS)
\\Porto Alegre, Brazil\\
\small{\texttt{\{acazevedo,ejscheid,mffranco,granville\}@inf.ufrgs.br}}}
}



\maketitle

\begin{abstract}
SSL/TLS is a fundamental technology in the network protocol stack that enables encrypted data transmission and authentication of web domains. However, the current model relies on a small number of Certificate Authorities (CAs) to provide and validate certificates, thus creating a highly centralized ecosystem. In this paper, we analyze the degree of centralization of certificate provisioning from CAs in two major political groups: Brazil, Russia, India, China, and South Africa
(BRICS) and the European Union (EU). We have found that over 75\% of certificates for both BRICS and EU domains originate from CAs based in the United States, indicating possible risks to their digital sovereignty due to the high level of external dependency. This indicates the need for nations within those groups to research alternatives to reduce the high level of dependency on foreign CAs and increase their digital autonomy.

\end{abstract}

\begin{IEEEkeywords}
Digital Sovereignty, Network Analysis, SSL/TLS, Network Measurement, Digital Certificates
\end{IEEEkeywords}

\section{Introduction}
\label{sec:intro}

The increasing centralization of services and infrastructure underlying the digital realm has become a topic of growing interest in recent years, due to the possible impacts that such a characteristic can bring upon the digital sovereignty of nations~\cite{digitalsov}. This discussion gained traction in the political sphere in the European Union (EU) and the intergovernmental organization of Brazil, Russia, India, China, and South Africa (BRICS), especially after the case of Eduard Snowden, with the leakage of information reporting a mass surveillance program from the United States, leading to governmental initiatives to protect data from its citizens from external actors and ensure~privacy~\cite{afterSnowden}. 

Digital sovereignty relates to several layers, ranging from subjects such as data regulation, portability, and transfer to geopolitical aspects, including trade dependence and sustainability~\cite{roleOfNetCentralization}. Network sovereignty is one of its pillars, which may include elements of traffic centralization, interoperability, neutrality of communication, hosting of critical services, and dependence of manufacturers in infrastructure~\cite{janardhanan2024networksovereigntynovelmetric}.



Secure Socket Layer (SSL), a security protocol that was designed to ensure privacy and encryption in network communication, is one of the key technologies behind services provided to end-users in the layers of the conceptual framework of network sovereignty in the OSI model, as depicted in~\cite{roleOfNetCentralization}. It was developed in the early 1990s and evolved into a more robust protocol known as Transport Security Layer (TLS), which fixes some of the existing vulnerabilities in SSL protocol while maintaining key features such as encrypted communication channels~\cite{tls}. 
Most popular web browsers only recognize a select amount of Certificate Authorities (CAs), leading to a concentration of certificate provisioning in the hands of a few responsible companies. This trait has also increased in the last few years, with a reduction in the mean chain of trust length, suggesting that certificate provisioning is becoming more dependent on fewer CAs~\cite{longitudinalCertificate}.

Disruption of the chain of trust from SSL/TLS CAs can lead to an outage of important and popular websites, as well as critical electronic-Government (e-Gov) services. For instance, the ongoing war between Russia and Ukraine provoked sanctions by Western countries against the Russian industry. As a result, there was a mass SSL/TLS certificate revocation for Russian domains from CAs. This caused the Russian government to issue TLS certificates through their own CA, though most popular web browsers used in the West, such as Google Chrome and Mozilla Firefox, still do not recognize it as trustworthy, requiring Russian citizens to adopt different browsers and for Russian domains to migrate to the government-provisioned~CA~\cite{russianTLS}. 

This depicts the possible impacts that this disruption could bring upon the digital sovereignty of a nation. Thus, measuring and understanding the influence that the centralization of infrastructure behind certificate provisioning can cause is fundamental. Therefore, in this paper, we propose an approach for the analysis of SSL certificates to verify the level of centralization of CAs for certificate provisioning, in order to evaluate its possible impacts on digital sovereignty and provide insights to foster the discussion on network sovereignty based on the SSL context.

The main contributions of this paper are as follows: 

\begin{itemize}
  \item Present an approach to analyze the centralization of SSL/TLS certificate provision;
  \item Provide a solution to gather information from certificate chains of domains belonging to a specific region;
  \item Using the presented solution, analyze the current scenario of SSL/TLS certificate provision centralization in countries from EU and BRICS using the proposed metric.
\end{itemize}

The remainder of this paper is structured as follows. In Section~\ref{sec:back}, we provide an overview of background knowledge and examine related work on digital sovereignty and SSL/TLS. In Section~\ref{sec:mesapp}, we define our approach to collect data and analyze whether there is centralization in SSL/TLS certificate provisioning, with details on the implementation. Section~\ref{sec:eval} presents our evaluation and analysis of the current scenario, followed by Section~\ref{sec:disc}, where we discuss the key findings of our approach. Finally, in Section~\ref{sec:conc}, we present our conclusions and possibilities for future work on the topic.

\section{Background and Related Work}
\label{sec:back}


This section describes the concept of network sovereignty, outlines current research efforts on the topic, and explores the existing literature related to the analysis of transport and session layer security, specifically TLS and SSL.

\subsection{Network Sovereignty}

With the advance of technology and the popularization of web services, the concept of digital sovereignty has also evolved over the years. Early notions were linked to developing the national technology industry to avoid external dependencies and become self-sufficient~\cite{whatSovMeansInDigital}. Although this idea is still present, it has expanded to encompass the elements and characteristics of the complex infrastructure behind the digital realm, such as data flows, privacy, security, and communication protocols~\cite{roleOfNetCentralization}. In this sense, the discussion on digital sovereignty became popular due to several data privacy and security incidents, such as the report of the mass surveillance program by the United States (US)~\cite{185057}. Hence, this boosted the political discourse on the EU and BRICS to perform legal changes ensuring digital autonomy and the protection of national data~\cite{belli2021brics, fightForDigSovEU}.

Key challenges to advance in this topic on political discussions have been the lack of metrics to analyze the level of digital sovereignty of a country in certain aspects, and the unclear definition of the entire stack that composes digital sovereignty. Most of the current literature addresses the term data sovereignty in different aspects such as cloud~\cite{cloudservice}, data geolocation~\cite{dataSovReview}, and information systems~\cite{vonScherenberg2024}, not covering other terms, such as network sovereignty and internet sovereignty. Hence, resulting in a partial view of the full scope of digital sovereignty. 

One of the key aspects that constitute the digital infrastructure is the network infrastructure and its communication protocols~\cite{techSovMissingThePoint}. They are related to the control and governance of data flows, thus being fundamental to the topic of digital sovereignty as they dictate the autonomy over national data and critical services. \cite{janardhanan2024networksovereigntynovelmetric} discusses how network sovereignty is related to the dependency of single manufacturers, as simultaneous failures on them could cause disruption of the network. It also showcases the difficulty in establishing a sovereign network due to the lack of metrics for measuring the level of sovereignty of a network and because of geopolitical influences. Thus, it proposes a metric for evaluating network sovereignty based on the diversity of manufacturers involved in the network path between source and destination addresses. Yet, only a few works such as~\cite{roleOfNetCentralization,noms24DNS} explore the network protocol stack and investigate their possible impacts on digital sovereignty, such as the DNS protocol. It investigates and maps internet domains names to their Authorative Name Servers (NSes) and the organizations behind them, determining the country/region of the Autonomous Systems (AS) that operates those NSes, with results showing, among other findings, that most of the top-10 DNS providers are from the US. This brings attention to the necessity of further investigation on the topic, addressing other network protocols for analysis.

\subsection{Transport and Session Layer Security}

SSL/TLS ensures secure and encrypted data transmission, being fundamental to several services such as online banking, Electronic e-Gov services and secure email transmission. \cite{8249081} analyzes possible security issues due to the dependency on the current model of CAs, highlighting issues within the model design and implementation that relate to certificate revocation and governance of CAs. It also brings attention to possible security breaches, as those CAs are often a target for attacks, leading to significant risk.~\cite{trustMeRootCA} explains how the mechanisms of authentication for SSL/TLS based on this CA model depends on the trustworthiness of a CA provider, and addresses possible issues based on the dependency of these companies and government institutions for attesting the identity of SSL service providers. The work also correlates the country of origin of analyzed CAs with different indexes, such as the Corruption Perception Index, Freedom on the Net Index, World Press Freedom Index and Legal Status of Capital Punishment, with results showing a varying level on those indexes, leading to questions on the trustworthiness of CAs associated with those countries.

\cite{blockchainSSL} proposes an alternative using a blockchain-based model for certificate and revocation transparency while also increasing security. SSL/TLS web servers would cooperate by publishing their certificates with CAs in the global certificate chain, thus providing a greater transparency of certificate validation, since a certificate would need to be validated by a majority of trusted nodes in order to be accepted in the certificate chain. \cite{governmentSSLAttack} discusses a possible attack on SSL where government authorities could compel CAs to emit false certificates leading to unsecured and unencrypted communication. It also emphasizes that websites should consider the country of origin of the CA that they use, especially when they are different from the country where user data is held, as it could expose this data to compelled certificate attacks. Thus, it highlights the possible impacts that depending heavily on foreign CAs can have. \cite{sslInvalidCertsMeasure} performs a study of invalid SSL certificates across the web, showing that most of them origin from end-user devices that reissue new invalid certificates. It also presents a technique that shows that they could be used to track end-user devices. Although it sheds light to the number of invalid certificates present on the internet ecosystem, it does not perform an analysis on the possible impacts that they present to the digital sovereignty of a nation. 

In the current literature, there seems to be a gap in the research on the level of dependency on foreign CAs for certificate provisioning and mapping its possible impacts on digital sovereignty. Such dependency poses a risk to a nation's digital autonomy since certificates can be revoked due to sanctions or political pressure from other countries, or even compelled by external governments. Our approach aims to fill the existing gap in the current literature by analyzing the centralization of CAs and certificate provisioning for SSL/TLS protocol, a key technology of the network stack, and the impacts of such characteristics in digital sovereignty.

\section{Measurement Approach}
\label{sec:mesapp}

\newcommand*\rot{\multicolumn{1}{R{45}{1em}}}

The approach is based on collecting popular Internet domains from countries belonging to the EU and BRICS and fetching their certificate chain to inspect characteristics such as \1 the organization that issued the certificates (i.e., the Certificate Authority (CA)), \2 the certificate issuers countries of origin and \3 both start and expiration date for each certificate. Table~\ref{tab:sample-data} presents sample data collected from a domain obtained from the list of popular internet domains.

\begin{table}[h]
  \centering
  \caption{Certificate Information Example}
  \label{tab:sample-data}
  \begin{tabular}{r|c}
  \hline
    \textbf{Data Type} & \textbf{Example} \\ \hline
    \textit{Domain} & \texttt{www.wikipedia.org} \\ \hline
    \textit{Subject} & \texttt{CN=*.wikipedia.org} \\ \hline
    \textit{Issuer} & \makecell{\texttt{CN=}DigiCert TLS Hybrid ECC SHA384 2020 CA1 \\ \texttt{O=}DigiCert Inc \\ \texttt{C=US}} \\ \hline
    \textit{Serial} & 7f2f35c87a877af7aefe947993525bd \\ \hline
    \textit{Not Before} & 2021-04-14 00:00:00 \\ \hline
    \textit{Not After} & 2031-04-13 23:59:59 \\ \hline
\end{tabular}
  
\end{table}

The subject field identifies the entity that owns the certificate. It can contain information such as the Common Name (CN), representing the domain name of the entity; Organization (O), to identify the organization that holds the certificate; and Country (C), for the country of the entity. The \textit{Issuer} field represents the CA that issued the certificate, and contains the same fields: CN, for the name of the CA; O, for the CA organization; and C, for the country of origin of that CA. \textit{Serial} is a unique identifier for the emitted certificate within the responsible CA, and it ensures that they can be tracked and revoked individually. \textit{Not Before} and \textit{Not After} fields define, respectively, the earliest date and time when the certificate becomes valid, and the latest date and time that a certificate should be considered valid. After this period, a certificate can no longer be considered as trusted.

\begin{figure}[ht]
    \centering
    \includegraphics[width=1\linewidth]{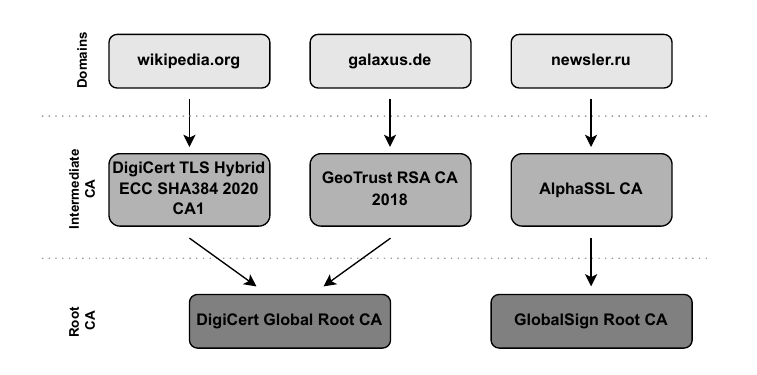}
    \caption{SSL/TLS Certificate Chain Example}
    \label{fig:sslchain-example}
\end{figure}

Figure~\ref{fig:sslchain-example} depicts a sample of the SSL/TLS certificate ecosystem using popular website domains.

Each HTTPS domain should present an SSL/TLS certificate, which is the leaf certificate. It is issued to a website through an Intermediate CA and contains information about the server's public key and the owner of the server. An Intermediate CA acts as a middleware between the domain and the Root CA, by signing certificates using its private key and establishing a hierarchy from the Root CA to the server. Hence, the Root CA is the highest level in the certificate chain. It signs certificates for Intermediate CAs, thus completing the chain of trust from a domain up to the Root CA. Therefore, it is possible to traverse the certificate chain from the leaf, present in the domain, to its Root CA.

\begin{figure}[ht]
    \centering
    \includegraphics[width=1\linewidth]{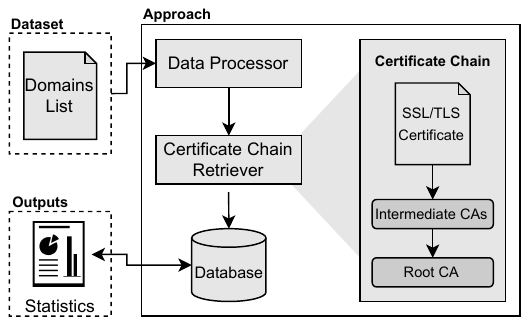}
    \caption{Overview of the Approach}
    \label{fig:approach}
\end{figure}

The defined approach presented in this section allows us to analyze the overall status of centralization of SSL/TLS certificate provision for domains in the political blocks of BRICS and EU, and to further investigate its possible impacts on the network sovereignty of the countries that compose them. Additionaly, it provides technical data to the geopolitical and economical discussion on digital sovereignty, thus facilitating further investigation on the topic. Figure~\ref{fig:approach} depicts the framework of the proposed methodology.

The first step of Figure~\ref{fig:approach} consists of obtaining a list containing popular domains from countries within those political blocks. These types of lists are commonly used in research to perform measurements of the web and to evaluate prototypes~\cite{topplingLists, Le_Pochat_2019}. The second step encompasses obtaining the chain of certificates from each domain in the list of popular websites and storing this data. As shown in Figure~\ref{fig:sslchain-example}, by obtaining the leaf certificate present in a website, it is possible to traverse the certificate chain going through intermediate CAs until reaching the root CA. After processing the list of domains and storing their certificate chains, statistics can be retrieved from the data obtained. 

The implementation of the proposed framework and the results of the evaluation performed are publicly available in an anonymous way at~\cite{githubSSL}. In our solution, we relied on the Chrome User Experience Report (CrUX) dataset, which contains the top million most popular websites reported by Google Chrome~\cite{crux}. Although there are other lists such as Alexa Top Million and Tranco Top Million, the CrUX list seems to be the most accurate for assessing web popularity~\cite{topplingLists}. In order to retrieve the certificate chain from domains present in the domain list dataset, our solution creates a socket connection to each one of them through the SSL Python library, which provides a wrapper for the OpenSSL system-native library, and allows us to access SSL/TLS encryption and certificate facilities. Hence, it is possible to fetch the certificate chain of website domains and store them in a CSV file for further analysis. A mechanism of retry is implemented, in case that the socket connection fails to be established or any errors occurs during the certificate retrieval. If the errors persist, then the domain gets excluded from the final database.
\section{Evaluation and Analysis}
\label{sec:eval}
The measurements were performed considering the CrUX dataset from February 2025. In total, we have analyzed approximately 232,490 domains from both BRICS and EU.


\subsection{Identifying Top Providers}

\begin{figure}[ht]
    \centering
    \includegraphics[width=0.95\linewidth]{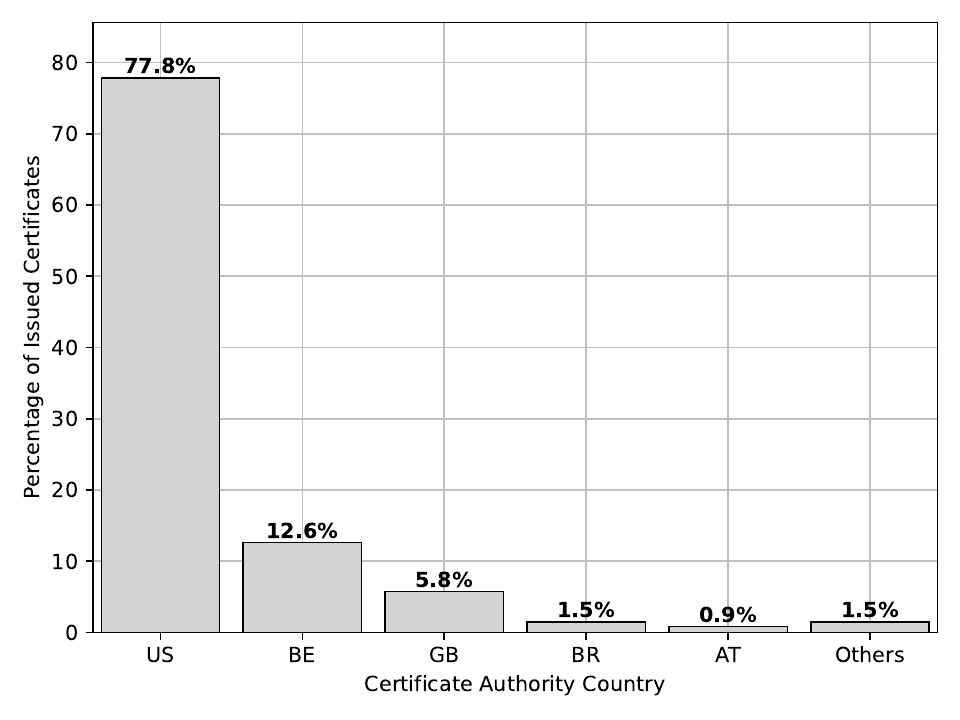}
    \caption{Top-5 Countries Issuing SSL Certificates for BRICS Domains}
    \label{fig:top_brics}
\end{figure}

\begin{figure}[ht]
    \centering
    \includegraphics[width=0.95\linewidth]{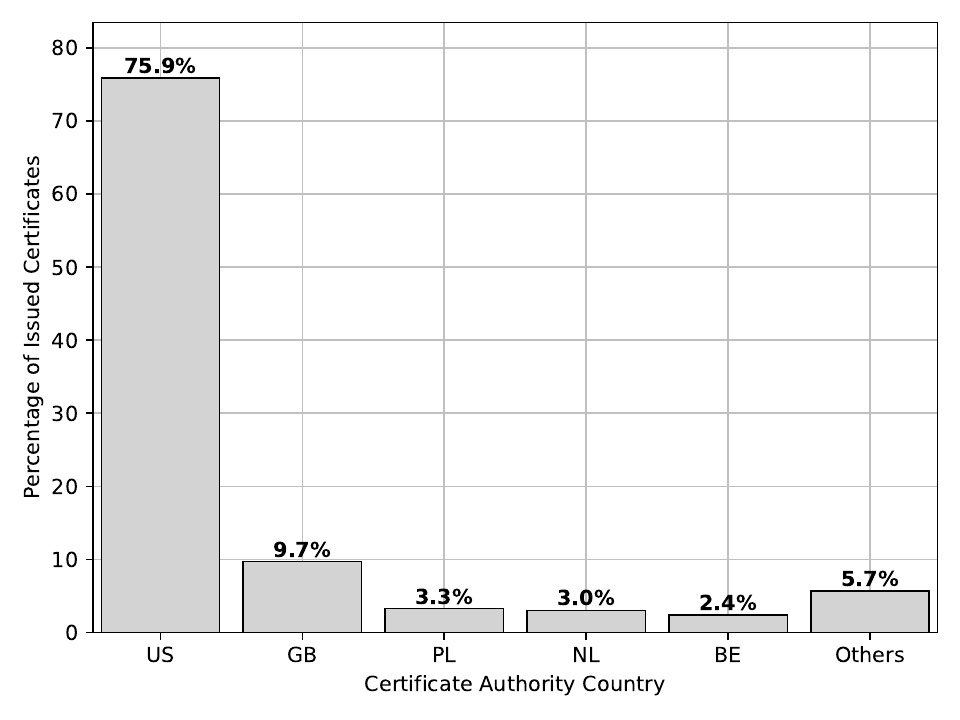}
    \caption{Top-5 Countries Issuing SSL Certificates for EU Domains}
    \label{fig:top_eu}
\end{figure}

Figures~\ref{fig:top_brics} and~\ref{fig:top_eu} demonstrate the Top-5 countries of CAs that issue SSL certificates for domains from BRICS and EU, respectively.

The results indicate an extremely high concentration of certificate provisioning for domains in both BRICS and the EU from CAs hosted in the United States (US), being responsible for 77.8\% and 75.9\%, respectively. This characteristic is in accordance with other protocols previously investigated in the literature, such as DNS providers~\cite{roleOfNetCentralization} that are also highly concentrated in the US.

In BRICS, the remaining 22.2\% is split between Belgium (12.6\%), Great Britain (5.8\%), Brazil (1.5\%), Austria (0.9\%), and other diverse countries with less significant portions. As for the EU, other major providers include Great Britain (9.7\%), Poland (3.3\%), Netherlands (3.0\%) and Belgium (2.4\%). This indicates a high level of external dependency, even though it is noticeable that the EU presents more in-group certificate emissions than BRICS.

Tables~\ref{tab:top_CAs_brics} and~\ref{tab:top_CAs_eu} provide an analysis of the Top-5 CAs issuing certificates for domains in BRICS and the EU, respectively. Each table contains the CA’s ranking position, total number of issued certificates, and country of origin, offering insights into the aspect of centralization of certificate provisioning within these regions.

\begin{table}[h]
    \centering
    \caption{Top-5 Certificate Authorities issuing SSL certificates for BRICS domains.}
    \label{tab:top_CAs_brics}
    \renewcommand{\arraystretch}{1.2}
    \begin{tabular}{cccc}
        \hline
        
        \textbf{Position} & \textbf{Certificate Authority} & \textbf{Certificates} & \textbf{Country} \\
        \hline
        1st & Let's Encrypt         & 31,848 & US \\
        2nd & Google Trust Services & 15,484 & US \\
        3rd & GlobalSign nv-sa      &  9,544 & BE \\
        4th & Sectigo Limited       &  4,336 & GB \\
        5th & DigiCert Inc          &  4,188 & US \\
        \hline
    \end{tabular}

\end{table}

\begin{table}[h]
    \centering
    \caption{Top-5 Certificate Authorities issuing SSL certificates for EU domains.}
    \label{tab:top_CAs_eu}
    \renewcommand{\arraystretch}{1.2}
    \begin{tabular}{cccc}
        \hline
        
        \textbf{Position} & \textbf{Certificate Authority} & \textbf{Certificates} & \textbf{Country} \\
        \hline
        1st & Let's Encrypt        & 57,539 & US \\
        2nd & Google Trust Services & 26,570 & US \\
        3rd & Sectigo Limited      & 13,297 & GB \\
        4th & DigiCert Inc         & 11,558 & US \\
        5th & Amazon               & 5,368 & US \\
        \hline
    \end{tabular}

\end{table}

In both BRICS and EU, four out of the top five CAs are based in the US, reinforcing the country's dominance in Public Key Infrastructure (PKI). \textit{Let's Encrypt} is the most dominant in both regions, possibly due to its characteristics, being a free CA that issues certificates with a 90-day validity period to encourage automated renewal practices~\cite{automatedSSL, letsencrypt}.

This raises concerns about the digital sovereignty of such regions, as disruption of services from these dominant CAs due to errors, revocation policies or even regulatory actions could affect a great amount of both BRICS and EU domains. 

\subsection{Analyzing Digital Sovereignty}

\begin{figure*}[ht]
\centering
  \begin{subfigure}[b]{.45\linewidth}
    \centering
    \includegraphics[width=1\textwidth]{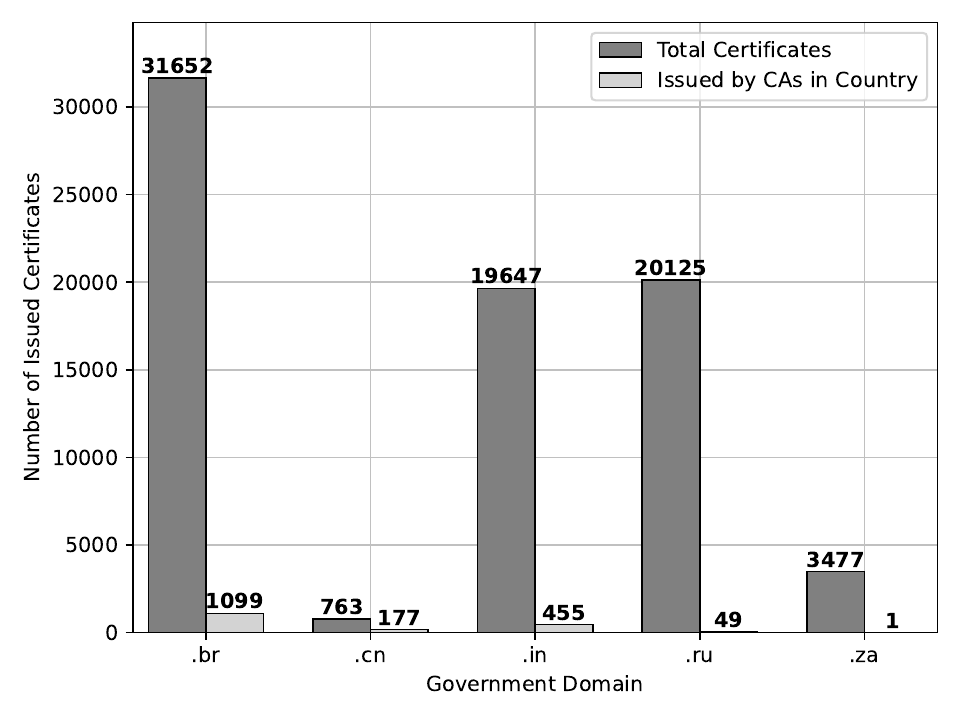}
    \caption{BRICS Countries}\label{fig:brics_certs}
  \end{subfigure}%
  \begin{subfigure}[b]{.45\linewidth}
    \centering
    \includegraphics[width=1\textwidth]{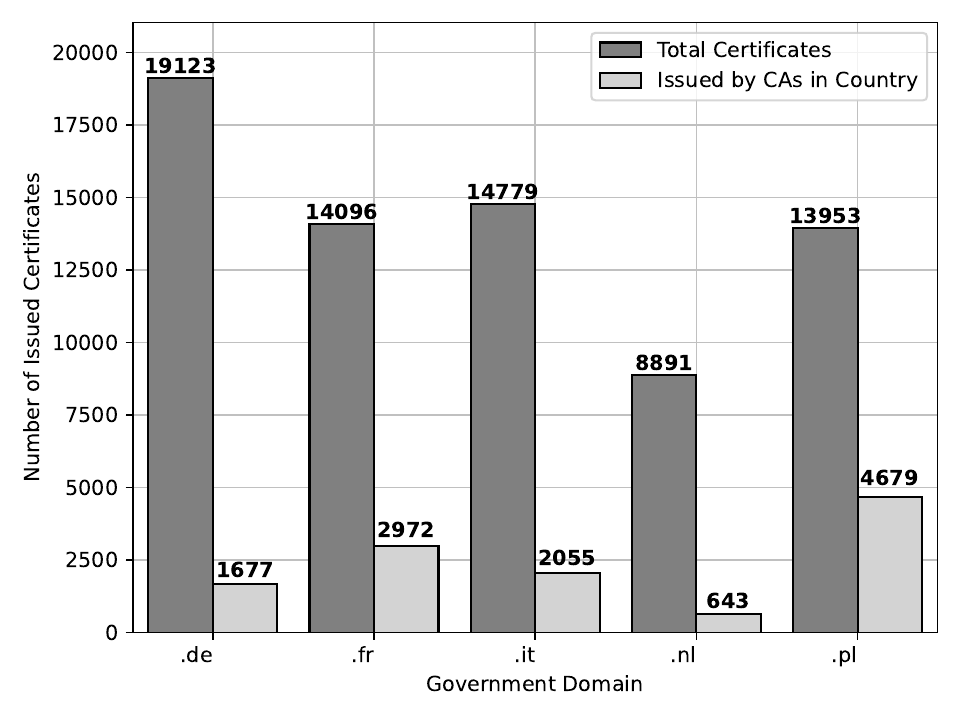}
    \caption{EU Countries}\label{fig:eu_certs}
  \end{subfigure}  \\ 
  \begin{subfigure}[b]{.45\linewidth}
    \centering
    \includegraphics[width=1\textwidth]{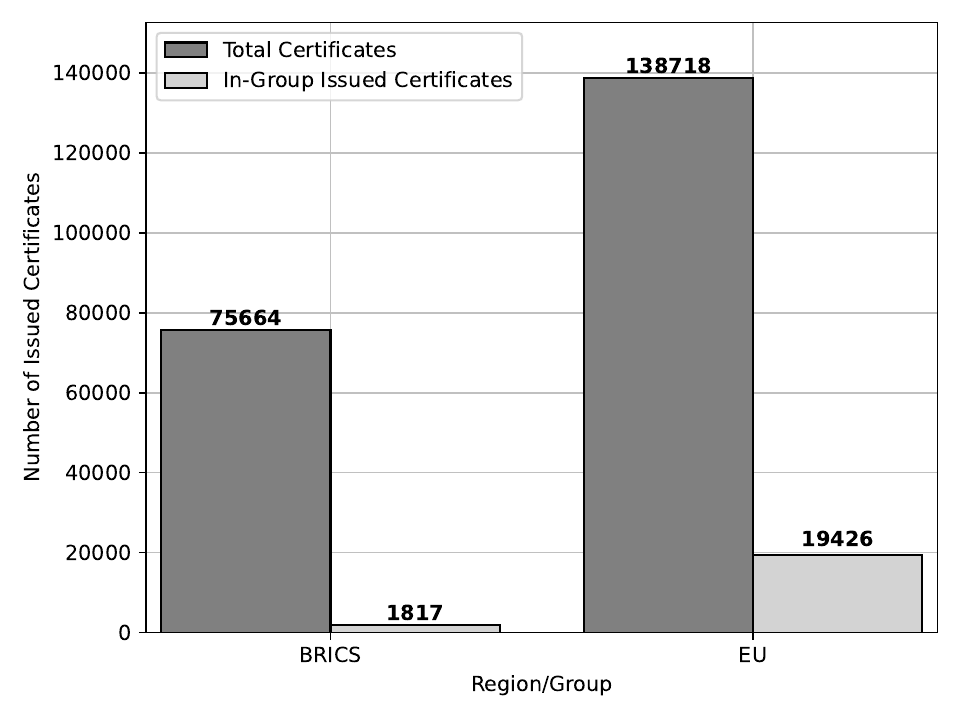}
    \caption{Per Group}\label{fig:groups_certs}
  \end{subfigure} 
  \caption{Certificate Issuing Distribution Separated by Countries and Group} \label{fig:brics-hosting}
\end{figure*}

Figures~\ref{fig:brics_certs}, ~\ref{fig:eu_certs} and~\ref{fig:groups_certs} provide more in-depth insight into the distribution of SSL certificate issuances within BRICS and EU countries, as well as in-group certificate issuance.

Brazil presents the most certificates among BRICS (31,652 certificates). Yet, only 1,099 were issued inside the country, representing 3.47\% of the total amount. This low in-country certificate issuing is seen other BRICS nations, such as India (2.32\%), Russia (0.24\%) and South Africa (0.02\%). The biggest discrepancy is China, with 23.20\% of the certificates from domains in the dataset being issued in the country.

Although there is a similarity in the aspect of external reliance for countries in the EU, it is possible to see that, for the top-5 countries with domains present in the dataset, they present a higher level of in-country certificate issuing. Poland and France shows the highest percentage, with 33.53\% and 21.08\%, respectively, while the lowest is in Netherlands with 7.24\%. This is also highlighted by Figure~\ref{fig:groups_certs}, which analyzes the overall issuance of certificates within these political groups.


Although BRICS has 1,817 internally issued certificates out of a total of 75,664, representing approximately 2.4\%, the EU demonstrates better internal certificate sovereignty. Within the EU, 19,426 certificates were issued internally out of 138,718, accounting for approximately 14\%.

\subsection{Certificate Validity Ranges}

\begin{figure}[ht]
    \centering
    \includegraphics[width=1\linewidth]{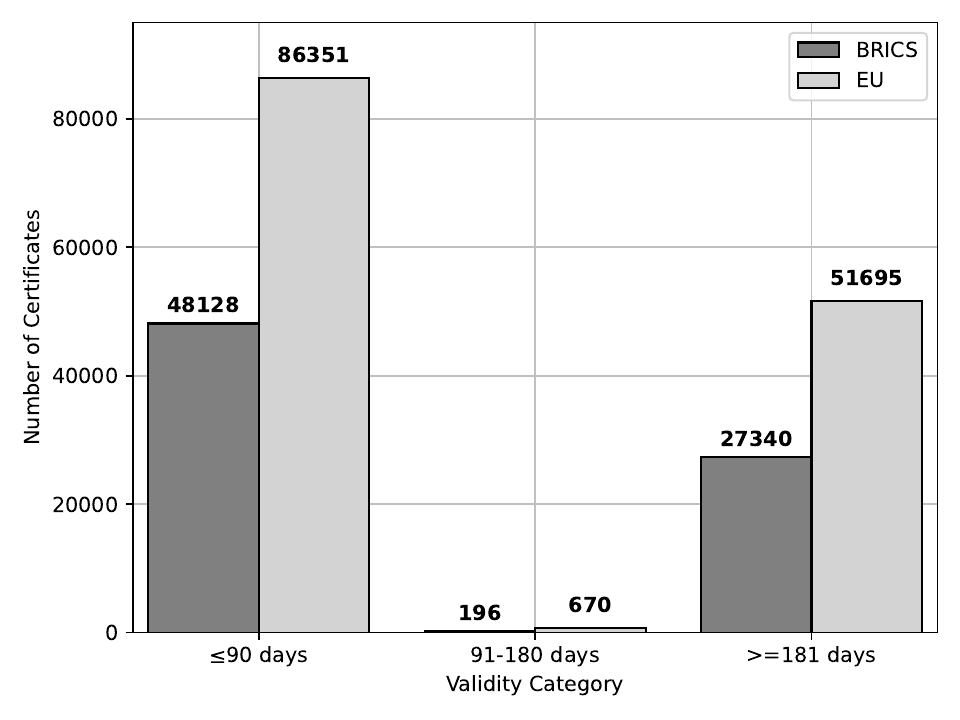}
    \caption{Certificate Validity Range in BRICS and EU}
    \label{fig:validity}
\end{figure}

In addition to the aspects of centralization of CAs, the validity period of certificates plays a key role in the availability, reliability, and security of domains. It directly impacts the frequency of the renewal process, the potential for disruption of services, and outdated data encryption, thus posing a concern to the digital sovereignty of a nation. Figure~\ref{fig:validity} analyzes certificate validity durations in BRICS and EU.

\begin{figure}[ht]
    \centering
    \includegraphics[width=1\linewidth]{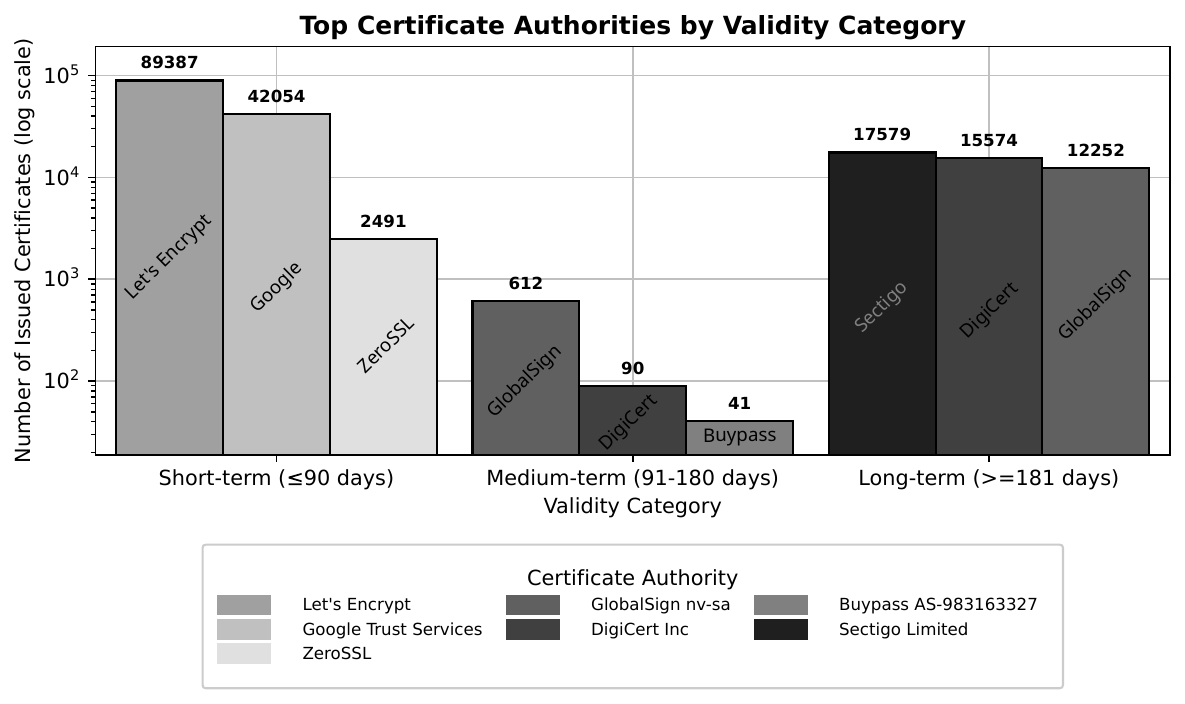}
    \caption{Certificate Validity Range in BRICS and EU}
    \label{fig:validity_ca}
\end{figure}

Both BRICS and EU show a higher prevalence of short-term certificates, with validation periods in a range of 90 days. They are responsible for 63.6\% and 62.25\% of the certificates within these groups, respectively. This has been an increasing pattern, possibly due to the high adoption of \textit{Let's Encrypt}, as shown in Tables~\ref{tab:top_CAs_brics} and~\ref{tab:top_CAs_eu}. Other large organizations may still issue long-lived certificates, likely including enterprise and government-issued certificates, as well as legacy systems that do not support automatic renewal, which explains the 36.1\% and 37.3\% of long-term certificates in BRICS and EU.
Figure~\ref{fig:validity_ca} depicts the current scenario of certificate issuing by company and validity range.

As discussed previously, Let's Encrypt seems to be the leader in certificate issuing for the range of 90-day certificates, followed by Google Trust Services and ZeroSSL. As they are free of fees and offer an automated renewal process, they are more accessible and thus widely popular than other CAs across the internet.

Medium-term certificates are very unpopular, in contrast to short-term and long-term certificates. For the latter, it is possible to notice a greater diversity in the distribution of CAs in certificate issuing. These companies often offer enterprise-grade certificates, with additional security features, which could explain its adoption, especially on big enterprises and government sectors. In such contexts, there might be regulatory and compliance requirements, in addition to complex IT infrastructure, that difficult to manage the usage of short-term certificates. Even though they are associated to higher security levels for their cryptographic key changes due to the frequent rotation of certificates, they can also pose a risk the the availability of services in case of downtime, endangering operational reliability. Further, Figure~\ref{fig:validity_ca} reinforces that the certificate ecosystem is dominated by a few key CAs. In total, 180,080 certificates are represented by 7 unique CAs, out of 214,282 collected certificates, representing 84\% of the total amount, thus indicating a clear over-centralized environment for certificate issuing.

Overall, the findings presented in this section highlight the high level of reliance on a small set of major CAs, that are mostly based in the US. The lack of in-group and in-country-based CAs points to a possible vulnerability in digital infrastructure stability. In this evaluation, we have focused our analysis on first-level certificates, \ie the leaf certificate emitted for the domains and their CA. Thus, we have not thoroughly investigated the characteristics of intermediate and root CAs, leaving it as an opportunity to extend this work in the future.
\section{Discussions and Key Observations}
\label{sec:disc}
Several insights can be obtained from the experiments performed in this work. It was possible to observe that SSL/TLS certificate provisioning is highly centralized, with most certificates in both BRICS and EU being issued by CAs based in the US.

\subsection{Impacts of SSL/TLS Centralization}

Although there is a clear sign of centralization in certificate provisioning, this characteristic is not inherently bad. Most popular browsers, such as Mozilla's, have a strict policy of security audits to validate and recognize CAs~\cite{mozillaCAprogram}. Moreover, trusting a malicious CA could cause severe impacts on the Web PKI, allowing the possibility of intercepting communication between a user and the web server~\cite{CAtrustManagement}. Popular major CAs, such as Let's Encrypt, tend to be in accordance with security protocols and best practices to ensure a high level of adoption and to be fully trusted by popular web browsers.

Another aspect is that, as depicted in Section~\ref{sec:eval}, the most popular CA in both BRICS and the EU is a company that provides free certificates with an automated renewal process. Other large commercial CAs usually charge fees for issuing premium certificates, often used in websites that contain sensitive data or payment systems due to their increased security. Further, nations may lack the necessary means to invest in or build their own secure and widely accepted CA infrastructure, thus leading to this external dependency on foreign CAs due to economic reasons.

Yet, this dependency on a small set of CAs increases the chances of geopolitical risks. As presented in Section~\ref{sec:intro}, Western sanctions on Russia led to a mass revocation of SSL/TLS certificates provisioned by US and EU CAs for Russian domains, leading to the creation of the Russian national CA~\cite{russianTLS}. This shows how the dominance of CAs controlling SSL/TLS infrastructure can cause political and economic impacts. Another example is the case of \textit{DigiNotar}, a popular Dutch CA that suffered a security breach, allowing attackers to perform Man-in-the-Middle attacks, a popular form of communication interception in SSL/TLS protocol~\cite{digiNotar}. This led the Dutch government, including governmental services, to find alternative CAs, as popular web browsers ceased to trust DigiNotar. Thus, relying heavily on a small set of CAs subjected to the possibility of cybersecurity threats can endanger the security and sovereignty of a nation.

Such cases also emphasize the roles of major technology companies responsible for popular web browsers in determining trustworthy CAs. In the case of national or regional CAs being established, they are still dependent on the recognition and acceptance of these companies. Consequently, nations are highly dependent on decisions made by foreign companies to be accessible and connected to the Internet.

\subsection{Concerns and Alternatives}

With the increasing political tension between the US, EU, and nations in BRICS, more cases of SSL/TLS certificate revocation could be provoked by sanctions. Although countries could try to incentivize the creation of their own national/regional CA, they are still dependent on the trust of popular web browsers, as depicted in the Russian case. Therefore, this alternative could lead to the isolation of such countries or regions from the global web.

Recent literature on the topic suggests that a possible alternative could be the use of blockchain-based PKI systems. This would lead to a decentralized trust system, distributed in multiple nodes, instead of concentrating it in a small set of CAs, reducing the risks of sanction-based certificate revocation and security breaches~\cite{blockchainPKI, blockchainSSL}. Although it could improve such aspects, it would also require adopting such systems by existing services and web browsers, which can be difficult due to technical and economic challenges.

Thus, the risks highlighted by the findings in this work need to be addressed while also considering the dual nature of SSL/TLS certificate provisioning centralization. Strategies such as regional CAs that meet the required criteria to be trusted by popular web browsers can enhance the resilience of the digital infrastructure in BRICS and EU. Still, they are not excluded from the possibility of distrust from such web browsers. A long-term strategy is required to ensure stability and security over key services and systems that rely on SSL/TLS protocol.

\subsection{Limitations}

The usage of the CrUX dataset has been proven accurate to represent popular domains in recent work~\cite{crux}. Yet, due to the limitations of China in terms of global internet access, it is likely that this impacts the list of popular domains. As a consequence, popular Chinese websites might not be accurately captured within CruX, causing a low number of Chinese domains in our dataset, as shown in our evaluation. As of February 2025, Chrome presents approximately 44\% of browser market share in China, being one of the most popular web browsers~\cite{chinaChrome}. Even though it is highly popular, there is still a low amount of statistics in these reports for Chinese domains. Future work could investigate if this aspect is also present in other datasets, such as the Tranco list.

It is also important to mention that while trying to retrieve the certificate chain from domains in the dataset, some failed and were not included in the final list. Most of the domains failed due to a \texttt{ECONNRESET} error. This occurs when the  TCP connection is closed from the server side due to one or more protocol errors. For instance, several errors were due to ``\textit{unsafe legacy renegotiation disabled}". This protocol error occurs when the server attempts to use an outdated and insecure TLS renegotiation method. Such feature is disabled by default in OpenSSL, due to its possible security impacts, as it allows Man-in-the-Middle attacks~\cite{cveMITM}. Another common protocol error faced during the obtention of certificate chain from domains in the list was ``\textit{unexpected EOF}". The possibilities for this issue range from a simple abrupt connection termination, to middlebox interference (e.g. firewalls on the server side) that deems the SSL/TLS handshake suspicious. Yet, the list of failed domains only represents, for BRICS and EU respectively, \textbf{9\%} and \textbf{5.2\%} of their total domains present in the list and thus does not invalidate the results and contributions provided in this work. In future work, the proposed framework can also be expanded to reduce the percentage of SSL/TLS handshake errors by implementing features that explore tha capabilities of OpenSSL library to overcome such issues, e.g. alternate the TLS version used to connect with servers, and disable weak cipher suites.

\section{Conclusion and Future Work}
\label{sec:conc}

This work has analyzed the SSL/TLS certificate provisioning centralization degree and its implications for digital sovereignty in BRICS and the EU. The findings show a significant reliance of nations within those groups on a small set of foreign Certificate Authorities (CAs), with over 75\% of certificates in both groups being issued by US CAs. We also show that, in comparison with each other, EU has a higher percentage of in-group issued certificates than BRICS. Additionally, our analysis also reveals that, even for the first level of certificates in the certificate chain, provisioning is concentrated in a few major companies in both groups, reinforcing the dominance of this small set of CAs in the Web PKI ecosystem. While this centralization is not inherently a bad trait, it can also pose substantial risks related to geopolitical control, economic dependence, and cybersecurity threats, directly impacting the digital sovereignty of nations.

The cases discussed in this work and the findings highlight that over-reliance on a small set of centralized CAs, especially in the cases of BRICS and the EU, can severely impact their key digital infrastructures and governmental services. Disruption provoked by security issues or political sanctions can lead to a mass revocation of certificates, causing accessibility issues and endangering national security. Regional CAs that comply with global security standards imposed by web browsers and organizations can increase the resilience of nations within such groups. However, they remain dependent on the recognition of popular web browsers, risking digital isolation. This highlights the challenges these groups face, which must try to reduce their dependability on external service providers while maintaining interoperability with the global web. Thus, alternatives should be researched for those groups to increase the resilience of their digital autonomy and decrease their dependency on foreign CAs.

Future work could expand the analysis to other levels of the certificate chain, thus investigating the current data in more depth and considering new datasets. Also, it is interesting to address other key network protocols and infrastructure to continue investigating their roles in the digital sovereignty of nations, to supplement the discourse on this topic further, and to guide it based on technical details. Moreover, this can provide a better understanding of how different layers of the network stack impact the digital sovereignty of nations and thus provide a foundation for political discussions on digital sovereignty and autonomy.



\bibliographystyle{IEEEtran}
\bibliography{bib/references.bib}
\footnotesize
All links were visited in March 2025.

\end{document}